# A Revised Incremental Conductance MPPT Algorithm for Solar PV Generation Systems

Meng Yue and Xiaoyu Wang
Sustainable Energy Technologies Department
Brookhaven National Laboratory
Upton, NY 11973, USA
yuemeng@bnl.gov, xywang@bnl.gov

*Abstract*—**A revised Incremental Conductance (IncCond) maximum power point tracking (MPPT) algorithm for PV generation systems is proposed in this paper. The commonly adopted traditional IncCond method uses a constant step size for voltage adjustment and is difficult to achieve both a good tracking performance and quick elimination of the oscillations, especially under the dramatic changes of the environment conditions. For the revised algorithm, the incremental voltage change step size is adaptively adjusted based on the slope of the power-voltage (*P-V*) curve. An accelerating factor and a decelerating factor are further applied to adjust the voltage step change considering whether the sign of the *P-V* curve slope remains the same or not in a subsequent tracking step. In addition, the upper bound of the maximum voltage step change is also updated considering the information of sign changes. The revised MPPT algorithm can quickly track the maximum power points (MPPs) and remove the oscillation of the actual operation points around the real MPPs. The effectiveness of the revised algorithm is demonstrated using a simulation.**

## I. INTRODUCTION

As one of the most promising renewable energy technologies, the installed capacity of the solar photovoltaic (PV) generation has increased dramatically in recent years. Although the cost of PV generation continues to drop, the economic competitiveness of solar PV energy is still low compared to the traditional energy sources, even with various local and federal policy instruments [1]. While it is desirable to further lower the cost and increase the efficiency of solar energy systems including both the solar panels and the power electronic devices, increasing the efficiency of the installed PV energy systems by simply improving the existing control algorithms should also be pursued. One way of achieving this is to modify the existing MPPT algorithms to extract more solar energy under various environmental conditions.

Many different types of MPPT algorithms have been proposed in the literature. As summarized in [2], different algorithms have their own pros and cons, in terms of complexity, accuracy and convergence speed, etc. Among the commonly used algorithms, the hill-climbing/perturbation and observation (P&O) method [3-6] is easy to implement using either analog or digital circuits. It periodically perturbs either the duty ratio of the converter or the the PV array operating voltage even when the MPP is achieved. Further, the "true" MPPT cannot be achieved using the P&O method since the operation point of the PV system is oscillating around the MPP. Under the conditions of continuous fast changing irradiance, the operating point might continuously deviate from the MPPs and eventually the optimal operation points cannot be achieved at all. These issues degrade the performance of the solar generation system. The fractional open-circuit voltage (or short-circuit current) method [7-9] needs only to sense one voltage (or current) parameter to approximate the MPP by using empirical parameters. The major issue related to this method is that the PV circuit has to be periodically operated in open-circuit (or short-circuit) conditions and may have significant impact on the grid operation. Other types of algorithms, such as those based on fuzzy logic control and neural network may accurately track the MPPs under different environmental conditions [2]. The MPPT performance, however, is not guaranteed since they both rely heavily on the algorithm developers and/or a significant volume of field data under all kind of conditions for the design and implementation of such algorithms.

The IncCond method (see, e.g., [10-17]) appears to be the most popular one in practice due to its medium complexity and the relatively good tracking performance. One of the major difficulties implementing the IncCond method is the selection of the (fixed) voltage change step size for simultaneously satisfying the tracking speed and maintaining the MPP. A large step size of voltage change helps the system rapidly approach the MPPs. On the other hand, this large value generally induces persisting oscillations around the MPP if no other special countermeasures were taken. The issues with using a small step size of voltage change are the opposite.

A simple and effective revised IncCond algorithm is proposed in this paper. An adaptive voltage step change scheme is first adopted based on the slope where the operation point locates on the *P-V* curve. An accelerating factor and a decelerating factor are then applied to further adjust the voltage step change considering whether the sign of the *P-V* curve slope remains the same or changes in a subsequent tracking step. The same information of sign changes is also used to update the upper bound of the maximum voltage step change. The adaptive voltage step change enables the PV system to quickly track the environment condition variations, i.e. reach and stay at the MPPs. In this way, more solar energy generation can be harvested from the PV energy systems. These improvements enable the quick response to the environment condition changes and rapid landing on the MPP. The revised method is easy to implement since it does not require knowledge of the *I-V* characteristics of specific PV panels and the parameters are easy to tune.

The revised IncCond algorithm is described in detail in Section II with an overview of various modified IncCond methods. Modeling of generic PV generation systems is presented in Section III for simulation purposes. Simulation results using the proposed MPPT algorithm will be shown in Section IV and concluding remarks are given in Section V.

## II. A REVISED INCCOND MPPT METHOD

The MPP is achieved by adjusting the terminal output voltage of a solar array through controlling the converter duty ratio. While the cell temperature can be easily measured, the irradiance is difficult to measure accurately, and the desired voltage at the MPP is hard to know exactly. Therefore, a test condition needs to be developed in order to determine whether the current operating point is the MPP or not without measuring the temperature and irradiance. For a solar panel, there is only one maximum power point for a given irradiance level and cell temperature. Note, the presence of a partial shading condition of a panel may cause multiple local maxima and is not considered in this paper, although the revised algorithm can be used together with the two-staged methods proposed in [16, 17].

The IncCond method uses the information of the solar *P-V* curve, i.e., at the left hand side of the MPP the slope is greater than zero, at the right hand side of the MPP the slope is less than zero, and the slope is zero exactly at the MPP. Therefore, the solar array terminal voltage needs to be increased when the slope is positive and decreased when the slope is negative. The slope $dP/dV$ can be calculated as,

$$\frac{dP}{dV} = \frac{d(IV)}{dV} = I + V\frac{dI}{dV} \quad (1)$$

with $dI/dV \approx \Delta I/\Delta V$ (i.e., the incremental conductance) in an implementation. Under the MPP condition, i.e., $dP/dV = 0$, the following relationship holds,

$$\frac{\Delta I}{\Delta V} = -\frac{I}{V} \quad (2)$$

The major difficulty with the IncCond method is the selection of the incremental step size of the duty ratio for adjusting the solar terminal output voltage. A fixed incremental step size of the duty ratio in general will not bring the array to the MPP because the operating point will oscillate around the MPP, i.e., either on the left or the right of the MPP.

Ref. [14] divided the entire *I-V* (current-voltage) curves into two domains using "square root" functions with all of the MPPs contained in only one of them. Therefore, the first step of performing MPPT is to bring the operating point to the domain that contains all of the MPPs. This method, however, requires a good understanding of the PV panel *I-V* characteristics that are panel specific. In [15], a so-called "Van Allen's oscillator" was added between the solar panel and the inverter for a purpose of balancing the power source and the load that continues changing. A simple proportional integral (PI) controller was developed to track the MPPs based on this configuration. It is intuitively easy to avoid a fixed voltage change step size by adjusting the increment proportionally to the steepness of the slope and eventually the increment of the duty ratio will become zero at the MPP, where the slope is zero, similar to the PI controller proposed in [15]. The implementation, however, appears to be very difficult because (1) the *P-V* curve steepness around the MPP can be very different for different operating conditions (i.e., the *P-V* curve for a lower irradiance level can be more flat) and (2) a sudden change in the operating condition of the solar array may produce a very large numerical difference in calculating the slope when the change occurs. Since the duty ratio is between 0 and 1, this may cause unacceptable change in the solar terminal output voltage and make it very difficult to bring the voltage back to normal. Note, refs [16] and [17] proposed two-stage methods mainly to avoid the local maxima caused by the non-uniform insolation experienced by the solar panels. The traditional IncCond method was still used after bringing the operating point close to the global MPP by using, e.g., monitoring cells in [17].

In this section, a simple and effective modified IncCond method is proposed based on observations that (1) in two consecutive tracking steps, a changing in sign of the slope (i.e., from positive to negative or from negative to positive) indicates that the increment step size is too large (otherwise, it may land on the MPP or the same side on the *P-V* curve) of the duty ratio; and (2) the same sign of the slope in two consecutive tracking steps indicates that the increment step size is too small (otherwise, the operating point may land on the MPP or the other side on the *P-V* curve). Based on these observations, the strategy proposed here is to (1) adjust the incremental step size considering the steepness of the slope; and (2) further adjust the incremental step size comparing the sign of slopes in two consecutive tracking steps, i.e., decrease the incremental size in the former case (e.g., by multiplying the incremental size by a factor *DEACC* such as 0.7) and to increase the incremental size in the latter case (e.g., by multiplying the increment by a factor *ACC* such as 1.2).

By applying this improved strategy, the solar array will approach the MPP in an accelerating manner after a change in the operating condition(s), and the magnitude of oscillation around the MPP may rapidly decrease until the test condition is considered to be satisfied. After landing onto the MPP, the duty ratio will not be adjusted until the operating condition changes again.

In the implementation, the upper- and lower-bounds for the incremental step size need to be defined to avoid extremely drastic changes in the duty ratio. However, the upper-bound is generally fixed and needs to be large enough to permit the rapid tracking of the MPP for a sudden change of the operation condition. The issue with a fixed upper bound is when the array starts tracking the new MPP and it quickly approaches the MPP and lands on the other side of the MPP, the duty ratio needs to be adjusted in the reverse direction using the incremental step, which could be large (due to the factor *ACC*) and remain large for some time (although the factor *DEACC* has been applied). At this point, having a large incremental step does not help because it may cause very large fluctuations or overshoot of the voltage before the MPP is achieved. Therefore, the second proposed improvement is, when the sign of the slope changes, the upper-bound is also decreased together with the incremental size. It is also preferred to maintain the upper-bound small nearby the MPP until the MPP is reached.

Note, in the implementation of the algorithm, a nominal incremental step size is pre-selected. After the test condition of the slope is considered to be satisfied, the duty ratio will not be changed but the incremental step size might become very small now, which, if not corrected, will cause the very slow response in the beginning of attempting to track the next MPP under a different operating condition. A simple solution is to reset the step size to the nominal value without adjusting the duty ratio when the MPP is considered to be reached.

Note also, the PV terminal output voltage may be very sensitive to the duty ratio, especially for duty ratio close to 0.0 or 1.0. The dc-dc converter input and output voltages should thus be selected such that the duty ratio is in the middle of the duty ratio range.

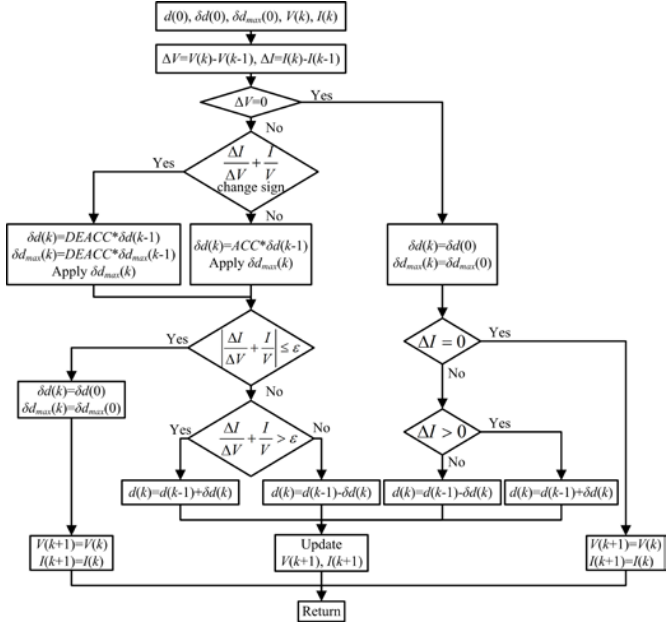

Fig. 1. Flowchart of the revised IncCond MPPT algorithm.

Based on the above discussions, the flow chart of the proposed modified IncCond algorithm is shown in Fig. 1. In Fig. 1, $V(\cdot)$ and $I(\cdot)$ are the terminal output voltage and current of the solar array, which will be adjusted according to the slope calculated by the dc-dc converter, $\delta d(k) = \delta d(k-1)$ $\times ACC \times |\frac{\Delta I}{\Delta V} + \frac{I}{V}|$ or $\delta d(k) = \delta d(k-1) \times DEACC \times |\frac{\Delta I}{\Delta V} + \frac{I}{V}|$ if the slope is not small enough (i.e., greater than a pre-selected constant $\varepsilon$ and equation (2) is still not satisfied), and the duty ratio $d(k) = d(k-1) + \delta d(k)$.

$\delta d(0)$ indicates the initial incremental size of the duty ratio, $\delta d_{max}(0)$ the initial upper bound of the incremental size. $\delta d(k)$ and $\delta d_{max}(k)$ are the updated (based on conditions discussed above, as indicated in Fig. 1) increment size and the upper boundary of the incremental size.

### III. MODELING OF PV ENERGY SYSTEMS

*A. Solar Array*

In general, a solar array consists of many solar modules connected in series and/or parallel, each module being manufactured by serially connecting a certain number of solar cells. A solar cell is essentially represented by an equivalent electrical circuit as shown in Fig. 2. For illustration purposes, modeling of a solar cell is briefly summarized in this part. Interested readers can find details in other references, e.g., [18].

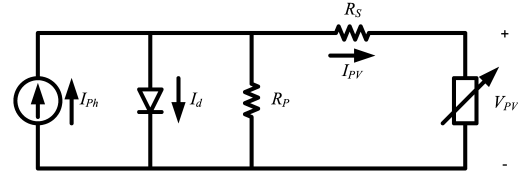

Fig. 2. An equivalent electrical circuit of a solar cell.

For the solar cell model in Fig. 2, the following equation can be derived,

$$I_{PV} = I_{Ph} - I_d - (V_{PV} + I_{PV} \times R_S)/R_P \qquad (3)$$

where $I_{PV}$ and $V_{PV}$ represent the solar cell terminal output current and voltage, respectively. $I_{Ph}$ is the photon current source, $I_d$ the diode current. The series resistance $R_S$ and shunt resistance $R_P$ are used to represent the power losses while the latter can generally be neglected. It is noted in equation (3) that the photon current and the diode current are temperature and irradiance dependent. For given panel temperature $T$ (Kelvin) and irradiance level $G$ ($W/m^2$), $I_{Ph}$ and $I_d$ can be calculated using the following equations,

$$\left.\begin{array}{l} I_{Ph}|_{T,G} = \dfrac{G}{G_{ref}} I_{Ph}|_{T,G_{ref}} = \dfrac{G}{G_{ref}} I_{SC}|_{T_{ref},G_{ref}} [1+\alpha(T-T_{ref})] \\[6pt] I_d|_{T,G} = I_0|_{T,G} \times [\exp(q(V+I\times R_S)/(nkT))-1] \\[6pt] I_0|_{T,G} = I_0|_{T_{ref},G} \times (\dfrac{T}{T_{ref}})^{\frac{3}{n}} \exp(\dfrac{-qE_g}{nk}(\dfrac{1}{T}-\dfrac{1}{T_{ref}})) \\[6pt] I_0|_{T_{ref},G} = I_{SC}|_{T_{ref},G_{ref}} /[\exp(qV_{OC_{ref}}/(nkT_{ref}))-1] \end{array}\right\}$$

$$(4)$$

In equation (4), $T_{ref}$ and $G_{ref}$ represent the reference cell temperature ($T_{ref} = 25°C$, i.e., 298 K) and reference irradiance ($G_{ref} = 1,000$ $W/m^2$) under the standard condition. $\alpha$ is the temperature coefficient and $I_{SC}$ is the short-circuit current of the solar cell. These are both constants that can be obtained from manufacturers' data sheets. $I_0|_{T,G}$ is the reverse saturation current of the diode, $n$ the diode ideality factor, $q = 1.602e^{-19}$ C the Coulomb constant, $k = 1.38e^{-23}$ J/K the Boltzmann constant. $E_g$ is band-energy gap (eV) and is given as $E_g=1.16-0.000702*T^2/(T-1108)$. $V_{OC}$ is the open-circuit voltage of the solar cell.

The series resistance can be solved using parameters at reference temperature and irradiance, i.e.,

$$R_S = -\frac{dV}{dI}\bigg|_{V_{OC_{ref}}} - nkT_{ref}/[I_0|_{T_{ref},G}\, q\exp(\frac{qV_{OC_{ref}}}{nkT_{ref}})] \quad (5)$$

where $\frac{dV}{dI}|_{V_{OC_{ref}}}$ can be obtained from the manufacturers' data sheet. After substituting the above parameters into equation (3), the I-V characteristics of the solar cell can be numerically computed for any given cell temperature and irradiance level and then used to represent a solar array consisting of interconnected modules and cells.

A buck-boost converter is used to step up the output dc voltage of a solar array such that a bulky step-up transformer can be avoided and perform the MPPT by controlling the duty ratio of the converter. See, e.g., [19] for more details.

## IV. SIMULATION RESULTS

The revised IncCond algorithm was implemented in an integrated Matlab-based power system simulation software EPTOOL that was developed based on the Power System Toolbox [20]. EPTOOL can be used to perform a transient analysis of the grid under faulted conditions and the solar irradiance and temperature changes. Only MPPT simulation results are presented here to validate the effectiveness of the revised algorithm using a hypothetical solar irradiance profile as the input to solar plant, as tabulated in Table I. The other input, the panel temperature is assumed constant of $25°C$ during the cloud transients.

A centralized solar PV plant consists of 17*170000 solar BP SX 150 panels [21]. The capacity of the PV plant is 433 MW or 4.33 pu (100 base MVA and under standard environmental conditions). A 50-machine-145-bus system was used as an example for carrying out the simulation purpose only.

TABLE I: VARIATION OF IRRADIANCE AT A SOLAR PLANT DURING A CLOUD TRANSIENT

| Time (s) | 0.0 | 0.2 | 0.7 | 0.9 | 1.2 |
|---|---|---|---|---|---|
| Irradiance ($W/m^2$) | 1000 | 20 | 200 | 300 | 400 |
| Time (s) | 1.5 | 1.9 | 2.5 | 3.0 | 4.0 |
| Irradiance ($W/m^2$) | 500 | 650 | 850 | 990 | 150 |
| Time (s) | 4.2 | 4.3 | 4.4 | 4.5 | 4.8 |
| Irradiance ($W/m^2$) | 120 | 20 | 210 | 330 | 340 |
| Time (s) | 4.9 | | | | |
| Irradiance ($W/m^2$) | 350 | | | | |

The conventional IncCond algorithm with a fixed incremental step size (0.001) of duty ratio is first applied and the MPPT is performed every 10 ms. Other parameters are selected as the following: $\delta d_{max} = 0.01$, and $\varepsilon = 5E-4$.

Simulation results for the solar array terminal output voltage and the deviation of the actual dc output power from the calculated maximum power points are shown in Fig. 3, from which one can observe the persisting oscillations around the MPPs in most of the time, i.e., the MPPs were not truly achieved. The reason for the oscillations is that, as implied in the algorithm description of Section II, the terminal voltage continues to be adjusted. Fig. 3 also indicates that the conventional IncCond algorithm is not able to track the MPP for a rapid variation of the irradiance since the output power deviations from the actual solar power generation are significant for the large change in irradiance at 0.2 s, although for slow variations it can provide acceptable performance. This significant power deficiency around 0.2 s is caused by an inability to adjust the panel voltage rapidly enough to compensate for the large decrement of irradiance level, as can be seen by comparing the top curves in Fig. 3 with those given in Fig. 4-5. This highlights the inefficiencies of selecting control parameters such as the incremental step size and the upper-bound in accordance with conventional MPPT algorithms. These inefficiencies related to the conventional IncCond method can be addressed by the modified algorithm proposed in this paper.

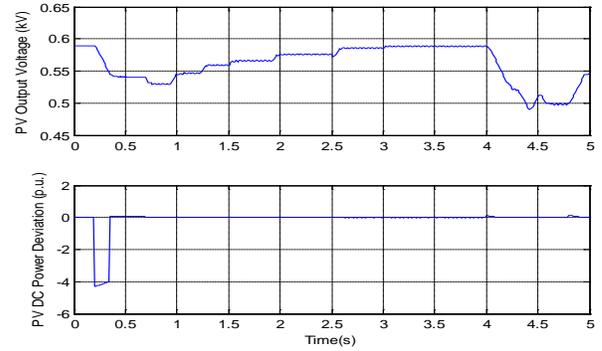

Fig. 3. Terminal voltage variation and deviated power output of the PV plant using conventional IncCond algorithm.

For the modified MPPT algorithm proposed in Section II, the deceleration and acceleration factors are applied in the modified IncCond algorithm with $DEACC = 0.8$ and $ACC = 1.2$ while other parameters remain the same. In the first scenario, the upper bound of the incremental step size of the duty ratio is fixed, i.e., $\delta d_{max}(k) = 0.01$, $k = 0, 1, …$. As shown in Fig. 4, the oscillations are eliminated quickly as the irradiance changes.

Fig. 4 also shows that the revised algorithm with the fixed upper-bound of the step size can quickly make the operating point reach the MPP, as the voltage level becomes quickly stable even after the sudden change of irradiance at time 0.2s. However, relatively large terminal voltage overshoot at changing points of irradiance is now introduced and must be addressed.

Fig. 5 shows further improvement of the tracking performance for a second scenario where an adaptive upper-bound of the incremental step is used. The overshoot at the change points of the irradiance have been significantly decreased. The output power deviation is also reduced.

Simulation also shows that the tracking performance is not sensitive to the associated parameters (e.g., $\delta d$), which makes parameter tuning very easy and the modified IncCond algorithm very robust.

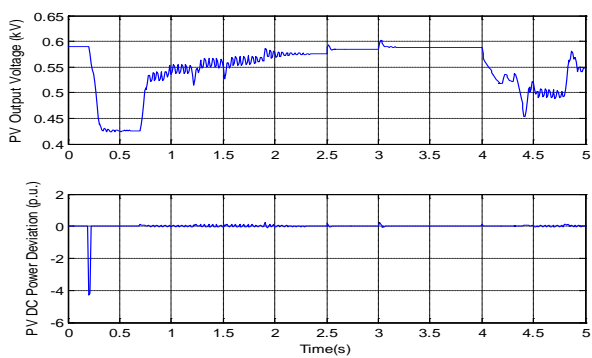

Fig. 4. Terminal voltage and deviated power output of the array using the modified IncCond algorithm (fixed upper bound of the incremental step size of duty ratio).

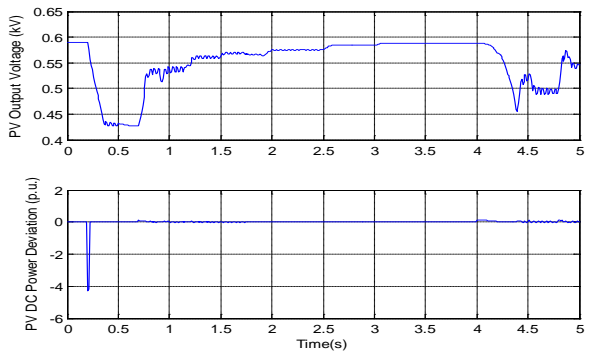

Fig. 5. Terminal voltage and deviated power output of the array using the modified IncCond algorithm (with adaptive upper bound of the incremental step size of duty ratio).

## V. CONCLUSIONS

A revised IncCond algorithm was presented in this paper for PV generation systems. Compared with traditional IncCond methods, the voltage step change is adaptively determined based on the slope of the *P-V* curve and the location of the operating points in two consecutive tracking steps such that the PV system can track the rapid change in environmental conditions while the oscillation of the PV system operating points around the MPP can be avoided. In addition, the upper bound of the voltage step change is assigned a factor, *DEACC* (less than 1) to constrain the step change when a change in the sign of slope is detected.. The simulation results demonstrate the effectiveness of the proposed algorithm. The robustness of the MPPT algorithm is also enhanced due to fact that the parameters can be easily tuned regardless of the PV systems and it does not require knowledge of the *I-V* characteristics of specific PV panels.